\documentclass[12pt,aps,floats]{revtex4}

\def\laq{\ \raise 0.4ex\hbox{$<$}\kern -0.8em\lower 0.62
ex\hbox{$\sim$}\ }
\def\gaq{\ \raise 0.4ex\hbox{$>$}\kern -0.7em\lower 0.62
ex\hbox{$\sim$}\ }

\def\N{${\cal N}$~}
\input epsf

\begin{document}

\title{The Shortest Scale of Quantum Field Theory }

\author{Ram Brustein, David Eichler, Stefano Foffa, David H. Oaknin}
\address{Department of Physics,
Ben-Gurion University, Beer-Sheva 84105,
Israel\\ email: ramyb,eichler,foffa,doaknin@bgumail.bgu.ac.il}


\begin{abstract}

It is suggested that the Minkowski vacuum of quantum field theories
of a large number of fields \N would be gravitationally unstable
due to strong vacuum energy fluctuations unless an \N
dependent sub-Planckian ultraviolet momentum cutoff is introduced.
We estimate this implied cutoff using an effective quantum theory of
massless fields that couple to semi-classical gravity and find it
(assuming  that the cosmological constant vanishes) 
to be bounded by $M_{\rm Planck}/{\cal N}^{1/4}$. 
Our bound can be made consistent with
entropy bounds and holography, but does not seem to be equivalent
to either, and it relaxes  but  does not eliminate the
implied bound on \N inherent in entropy bounds.

\end{abstract}

\maketitle

The existence of a fundamental value for the entropy of a black
hole (BH) \cite{BekHaw} which depends on a geometric property,
its area, seems startling since it appears to limit the number of
different types of elementary particles \N.  If there were a
sufficiently large number of particle species within a given mass
scale, then any black hole much larger than that scale could have
formed in a sufficiently large number of ways as to exceed any
bound on entropy which does not depend on \N. This argument goes
beyond holography \cite{holography,Bousso}, which states that the
entropy is established at the boundary, but which nevertheless
allows that surface-associated entropy to be proportional to \N.
In fact, according to  elegant arguments and calculations by
Bombelli {\it et al} and Srednicki\cite{Srednicki}, surface
entanglement entropy is indeed  proportional to \N
\cite{Susskind,Bek1}. But why then should \N  be limited in  any
field theory that is to be consistent with gravity? Moreover, if
gravity is the limit of a large $N$ gauge theory \cite{Maldacena}
how could it disallow large \N?

In an attempt to broach these issues, we show here that three
commonly made assumptions (i) that free, or weakly coupled quantum
field theories (QFT's) have empty Minkowski space-time as their
vacuum on scales greater than the familiar Planck length
$L_{p}$,(ii) that free, or weakly coupled QFT's can be used
to calculate the spectrum of quantum fluctuations in Minkowski
space-time background and (iii) that quantum fluctuations
gravitate as any other source of energy, are not compatible for
sufficiently large \N. Assumptions (i,ii) are commonly used to
calculate many physical observables, and have been tested
experimentally to very good accuracy at low
energy scales. Assumption (iii) has not
been tested experimentally, but is commonly believed to be
correct, and leads to the infamous cosmological constant problem
\cite{weinberg} due to the contribution of zero-point quantum
fluctuations to the energy density.

We propose that the resolution to the conflict  between
assumptions (i-iii) when the number of fields becomes large is
that quantum field theories which have flat space as their vacuum
have an \N-dependent limit on their shortest scale (ultraviolet
cutoff).  The logical possibilities are that either \N is limited,
that the scale of quantum gravity can be substantially lower than
Planckian, or that quantum fluctuations do not gravitate, making
assumption (iii) invalid. The latter possibility would also be a
possible solution to the cosmological constant problem.

We first show that the naive Minkowski vacuum of free QFT's of a
large number of fields can be gravitationally unstable. For such
theories, vacuum energy fluctuations in regions whose volume is
smaller than a certain ``critical" volume (which can, as large \N
is contemplated, be parametrically larger than a Planck volume)
would become so strong that they induce them to collapse. The Minkowski
vacuum then heads towards a BH slush unless some back reaction can
modify the naive Minkowski vacuum sufficiently to prevent this
final outcome. Since this can happen on scales parametrically
lower than Planckian for large \N, the problem is  not due to quantum
gravity effects, rather, classical gravity is amplifying energy
density seeds originating from field theoretic quantum
fluctuations, in analogy to  the germination of large scale
structure  during an inflationary phase of the early
universe.

We are not suggesting that the  vacuum actually  is unstable.
We merely point out that Minkowski space-time  would be
unstable if we try to extend field theory with a large number of
fields and semiclassical gravity beyond a certain scale. As for
what happens beyond that scale, perhaps the vacuum needs to be
redefined by non-perturbative effects due to 
 back-reaction on the vacuum (or the BH slush,
should it come to that). In any case, the final outcome represents
a significant modification of naive Minkowski space. While it may
remain Minkowskian at sufficiently large scales, it features an
ultraviolet cutoff that depends on \N, and thus restricts the
number of independent degrees of freedom available in a region of
given size. This number, when there is no cosmological constant,
generally increases as fractional power of \N, so it implies
neither an absolute entropy bound \cite{ebounds}, nor a linear dependence of
such a bound on \N, as in holography.

Energy fluctuations in the vacuum occur even though the vacuum is
an eigenstate of the total hamiltonian. For our concrete
discussion of energy fluctuations in the vacuum we consider a
free field theory of \N massless bosonic scalar fields, but our
results are applicable, with slight modifications, to physical
components of any kind of fields, such as fermions, gauge bosons,
etc.. We assume that the fields' masses are protected from
quantum corrections, for example, by supersymmetry, or a gauge
symmetry, so they are strictly massless. The restriction to
massless fields is mainly for convenience, allowing us to present
simpler analytic results which capture the essence of our
point.   We further assume that the cosmological constant has
been set to zero (at least to some accuracy), for example by
unbroken supersymmetry. The curvature of spacetime is therefore
much lower than Planckian, and we assume for simplicity that the
background space is Minkowski space. We do expect, however, that
the instability we will demonstrate persists in presence of a
(small) cosmological constant. We neglect possible renormalization of
Newton's constant, so that the scale of
quantum gravity is indeed the Planck scale. Throughout we emphasize
functional dependence on mass scales and \N (which we assume to
be a large parameter), and work in units in which $\hbar=c=1$.

The hamiltonian $H=\int d^3x\ {\cal H}(\vec{x})$, of a single
massless scalar field $\phi$ in Minkowski spacetime is given by
a volume integral over the hamiltonian density
${\cal H}(\vec{x})=
\frac{1}{2}\left[\left(\vec{\Pi}(\vec {x})\right)^2 +
\left(\vec{\nabla}\phi(\vec {x})\right)^2\right]\, ,
$
where $\vec{\Pi}$ is the momentum conjugate to $\phi$.
Separating space into two parts, an ``inside" region
of volume $V$ and an ``outside"
 region of volume $\widehat{V}$, the
total hamiltonian is simply given by
$H=H_V+H_{\widehat{V}}=
\int\limits_V d^3x\ {\cal H}(\vec{x})+
\int\limits_{\widehat{V}} d^3x\ {\cal H}(\vec{x})\, .
$

Although the vacuum state is an eigenstate of H, it is not an
eigenstate of
$H_V$ or $H_{\widehat{V}}$.
So in spite of the vacuum being an eigenstate
of the total Hamiltonian, the energy contained in the volume $V$
is subject to fluctuations, its dispersion given by
\begin{eqnarray}\label{DHV}
&& \left\langle\left(\Delta H_{V}\right)^2\right\rangle
=\frac{1}{8}\int\limits_V d^3y_1\ d^3y_2\ \Biggl[
\int\frac{d^3p\ d^3q}
{(2 \pi)^6} \times \Biggl\{
\nonumber\\
&& \ e^{\hbox{$-i(\vec{p}+\vec{q})\cdot(\vec{y}_1-\vec{y}_2)$}}
\left[pq + 2 \vec{p}\cdot\vec{q}+
\frac{(\vec{p}\cdot\vec{q})^2}{p q}\right]
\Biggr\}\Biggr]\, .
\end{eqnarray}
Note that if $V$ in (\ref{DHV}) is the whole of space, then the
integration over $\vec{y_1}$ and $\vec{y_2}$ produces a $\delta^3
(\vec{p}+\vec{q})$, forcing the momentum integral to vanish. The
dispersion of $H_{\widehat V}$, as expected, is equal to that of
$H_{V}$. This can verified by expressing $\int_{\widehat V}$ as
$\int_{I\! R^3} - \int_{ V}$ for the ${d^3x}$ and $d^3y$
integrations, and using the fact that each of the $\int_{I\! R^3}$
integrals gives a vanishing result due to the presence of
$\delta^3 (\vec{p}+\vec{q})$.

It is convenient to express (\ref{DHV}) as an integral of a
density, $ \left\langle\left(\Delta H_{V}\right)^2\right\rangle
=\int\limits_V d^3y_1 d^3y_2 F(|\vec{y}_1-\vec{y}_2|)\, , $ where
the density of energy fluctuations $F(|\vec{y}_1-\vec{y}_2|)$ is
given by the expression inside the square brackets on the r.h.s. of
(\ref{DHV}). Since $F$ depends only on
$x\equiv|\vec{y}_1-\vec{y}_2|$, we can perform all the integrals
in (\ref{DHV}), except for the $x$ integral, by using the equality
$ 1=\int_{0}^{\infty} dx \delta(|\vec{y}_1-\vec{y}_2|-x) $. The
result is the following convolution,
\begin{eqnarray}
\label{intx}
\left\langle\left(\Delta H_{V}\right)^2\right\rangle=
\int dx F(x) {\cal D}_{V}(x)\, ,
\end{eqnarray}
where the geometric factor ${\cal D}_{V}(x)$ depends only on
the shape of the volume $V$.

The energy dispersion calculation leading to
(\ref{DHV}),(\ref{intx}), has many similarities to the
calculation of the entanglement entropy of a subsystem of a pure
state \cite{Srednicki}. Since  the dispersion of $H_{V}$ is equal
to that of $H_{\widehat V}$ they can depend only on properties of
the common boundary of the two regions. This is perhaps
counterintuitive, one might have expected the dispersion of
$H_{V}$ to be extensive, proportional to the volume $V$, but, as
the previous argument shows, this is wrong. The fact that the
dispersion of  $H_{V}$ has to be a function of boundary
invariants and using dimensional analysis allows us to estimate
it in different setups. Of course, as it stands,
$\left\langle\left(\Delta H_{V}\right)^2\right\rangle$ is
ultraviolet divergent, being an operator of mass  dimension 2; to
define it we have to introduce an ultraviolet momentum cutoff
$\Lambda$. The exact form of implementing the cutoff will not
affect the nature of our results, but it will change details,
such as numerical coefficients of order of unity.

For the sake of concreteness and clarity, we restrict our
attention for the moment to the case of a spherical volume $V$ of
radius $R$. We expect similar results when different geometries
are considered, and present some examples later on. On
dimensional grounds, the energy dispersion in a sphere of radius
$R$, $\Delta E(\Lambda,R)=\sqrt{(\Delta H_{Sphere})^2}$, is given by
$\Delta E(\Lambda, R)={\cal E}(R \Lambda) \Lambda\,$. We now proceed to
find the analytical expression for the function ${\cal E}(R
\Lambda)$. The geometric factor for a spherical volume is given by
\begin{eqnarray}
\label{DR}
{\cal D}_{Sphere}(x, R)=
\frac{\pi^2}{3}x^2 (x - 2R)^2 (x + 4R) \quad 0<x<2R\, ,
\end{eqnarray}
and, of course, vanishes for $x>2R$. Since the density $F$ is
ultraviolet divergent, it has to be regularized. We implement a
particularly simple regularization procedure by inserting factors of
$e^{\hbox{$-p/\Lambda$}}$ and $e^{\hbox{$-q/\Lambda$}}$ which
suppress momenta larger than $\Lambda$ in the momentum integrals
of eq.~(\ref{DHV}).  Now we can explicitly evaluate $F$,
\begin{eqnarray}\label{FA1}
F(x,\Lambda)=\frac{\Lambda^8}{2\pi^4}\frac{3-10(\Lambda x)^2+
3(\Lambda x)^4} {\left(1+(\Lambda x)^2\right)^6}.
\end{eqnarray}
Notice that $F$ has an over all factor of $\Lambda^8$ as required
by its dimensionality, that the maximal value of $F$ is at zero
$F(0,1)=\frac{3}{2\pi^4} \sim 0.015$, and that for large $x$, $F$
is positive and decreases as $x^{-8}$. Using (\ref{DR}) and
(\ref{FA1}), integral (\ref{intx}) for $\Delta E$ can be
evaluated explicitly,
\begin{eqnarray}\label{DEA1}
\Delta E(\Lambda,R)=\frac{(\Lambda R)^3}{\pi}
\left[\frac{8\left[5 + 4(\Lambda R)^2\right]}
{15\left[1 + 4(\Lambda R)^2\right]^3}\right]^{1/2} \Lambda\, .
\end{eqnarray}
For regions of different shapes and different cutoff procedures
we expect similar results and indeed have found similar results.

In a theory of  a large number of fields ${\cal N}$, the energy
dispersion $(\Delta E_{\cal N}(\Lambda, R))^2= \left\langle\left(\Delta
H_{Sphere}\right)^2\right\rangle$ is  proportional to \N, since
the contribution of each field adds up linearly, so $\Delta
E_{\cal N}(\Lambda, R)$  is given by
\begin{equation}
\label{edisp}
\Delta E_{\cal N}(\Lambda, R)=\sqrt{\hbox{\N}}{\cal E}(R \Lambda) \Lambda\, ,
\end{equation}
where ${\cal E}(R \Lambda)$ can be read off (\ref{DEA1}).

The energy fluctuation $\Delta E_{\cal N}(\Lambda, R)$ differs from
the expectation value of vacuum energy $\langle{H}\rangle$ (the
cosmological constant). For example, bosonic and fermionic fields contribute
with different signs to $\langle{H}\rangle$, so an exact cancelation, as in a
supersymmetric theory, is possible.  But we have explicitly checked that bosonic
and fermionic contributions to the dispersion have the same sign, as expected,
and cancellation is not possible. In addition, their \N dependence is
different, $\Delta E_{\cal N}(\Lambda, R)$ being proportional to $\sqrt{{\cal
N}}$, while $\langle{H}\rangle$ is generically proportional to \N. Moreover,
since we are dealing with fluctuations, it is clear that $\Delta E_{\cal
N}(\Lambda, R)$ should not be considered as ordinary, classical energy, but
rather as a stochastic, fluctuating quantity, with a typical lifetime given by
$T_{f}\sim \pi/\Lambda$. This estimate is based on the fact that the dominant
contribution to the fluctuation is given by the high momentum modes, which have
an energy of order of the cutoff scale $\Lambda$.  If we consider only modes
with energy less than some maximal energy scale $\Lambda^{*}$
($\Lambda^{*}<\Lambda$), the energy fluctuation $\Delta E(\Lambda^{*},R)$ which
is a subdominant contribution to the total fluctuation $\Delta E(\Lambda,R)$)
has a longer lifetime $T^{*}_{f}\sim \pi/\Lambda^{*}>T_{f}$.

With this remarks in mind, we now show that unless the number of degrees of
freedom of QFT's is bounded their Minkowski vacuum is
gravitationally unstable. Let us consider a theory of ${\cal N}$
massless scalar fields in a classical spacetime background. If
spacetime curvature is smaller than Planckian, then according to
assumption (iii), the energy-momentum tensor of the QFT can be
consistently used as a source in the classical Einstein equations
for the metric.

When the expectation value of the energy-momentum tensor vanishes
(recall that we have assumed that the cosmological constant
vanishes), one must also consider its fluctuations as a stochastic
source in the Einstein equations (see, for example, \cite{stoc}).
Adopting this prescription, we consider the gravitational effects
of the energy fluctuation in a given volume, eq.~(\ref{edisp}),
and we immediately encounter a potential problem. When a typical
energy fluctuation is within its own Schwarzshild radius,
$2 G_{N}\Delta E_{\cal N}(\Lambda, R) \gaq R$,
a BH could be created ($G_N$ is Newton's constant).

That a fluctuation is within its own Schwarzschild radius is not
sufficient information to determine whether a BH would actually
be formed. An additional necessary condition is that the
the travel time of light through the collapsing region must be
comparable to the mean lifetime of the energy fluctuation itself.
This means that only energy fluctuations with a lifetime
$T^*_{f}>R$ can create a black hole in a region of size
$R$, so only modes with energy less than $\Lambda^{*}\sim\pi/R$
are relevant to this process, and therefore the previous condition for vacuum
instability should be refined as follows
\begin{eqnarray}\label{GDER}
2 G_{N}\Delta E_{\cal N}(\Lambda^{*}, R) \gaq R\, ,
\end{eqnarray}
which, together with (\ref{edisp}), implies that the condition is
\begin{eqnarray}\label{boundR}
R^2 \laq 2 \pi {\cal E}(\pi) \sqrt{\cal N} L_{p}^2\, ,
\end{eqnarray}
where $L_p=\sqrt{G_N}$ is the Planck length.

For a large enough \N, the size of created BH's is large in Planck units,
so the initial induced curvature by each one of them is
small. 

The main limitation to our calculation comes from the fact that we have
treated the BH's as classical objects; this is a consistent procedure as long
as the evaporation time by emitting Hawking radiation $t_{\rm ev}= \frac{5 (8
\pi)^4}{4 \pi^3}\frac{M^3 G_N^2} {\cal N}$ is longer than the characteristic
classical time scale $t_{\rm bh}=2 G_{N} M$.  By setting $2 G_{N} M=R$, one
finds that only BH's with $R>R_{c}\equiv\sqrt{\frac{{\cal N}}{640 \pi}} L_p$
can be treated classically.  Thus
the validity of the classical treatment requires that
the r.h.s. of eq.~(\ref{boundR}) must be
larger than $R_{c}^2$, and hence
${\cal N}\laq (1280)^2 \pi^4 {\cal E}^2(\pi)\sim .5 \times 10^7$ 
(which by eq.(\ref{boundR}) requires 
$R\laq \sqrt{2560 \pi^3} {\cal E}(\pi) L_{p}\sim 50 L_{p}$).
Since BH evaporation process is the inverse
of BH formation process (which we have considered to derive (\ref{boundR})),
their strength is determined by the same parameters and couplings.
Therefore it is not possible to tune some of the parameters
of the theory to avoid BH evaporation with the 
context of the argument. 
The relative strength of the BH formation and evaporation
processes becomes a detailed quantitative question, which we cannot
address using our
methods beyond what we have just discussed. To
determine more reliably the limitation imposed on our arguments
by the BH evaporation process, a better treatment of quantum gravitational
effects and their interplay with field theoretic effects is required.

If indeed BH's are created, they are created at a rate of about $\Lambda^{*}$
and at a density of about close packing, making the vacuum of the
theory very different than Minkowski space-time, in contradiction
with assumption (i).

The gravitational instability of flat space that we have noted 
can be avoided  if $\Lambda$, the field theory UV cutoff, 
is low enough. Since $\Lambda^{*}\sim \pi/R<\Lambda$,
we obtain the following bound on the ultraviolet cutoff of the
theory
\begin{equation}\label{lambda}
\label{lLlp} \Lambda\laq \frac{\alpha} {{\cal N}^{1/4}} M_p\, ,
\end{equation}
where $M_p=1/L_p$ is the Planck mass, and 
$\alpha=\sqrt{\pi/2{\cal E}(\pi)}\sim 2.9$. This bound is subject to
the validity conditions which we have discussed above.

We have found  that for a given $\Lambda$, treating \N as a
variable, there is a critical value $(M_p/\Lambda)^4$ 
above which the vacuum becomes
gravitationally unstable. Alternatively, if \N is fixed and we
treat $\Lambda$ as a variable, we find that $\Lambda$ cannot be
made larger than a certain critical value, which is parametrically
lower than the Planck scale. Thus a large number of massless
fields ${\cal N}\sim 10^4$  (about the number of massless modes in
some string theories) can be admitted in a field theory provided
that the ultraviolet cutoff of the theory is sufficiently below
the Planck scale. In the context of low energy effective field theories
of weakly coupled strings, the cutoff scale is determined by 
the string length.  However, according to our arguments, the magnitude of 
the cutoff scale may be set by considerations that do not seem to 
have a direct connection to the perturbative definition 
of the string length, rather by BH's that strings may form.

We  conclude that in order to avoid a ``granular collapse'' of
spacetime, the number of fundamental degrees of freedom has to be
restricted if not bounded.  Because our calculation assumes a flat
background, we have not explicitly derived the mechanism by which
the scale of QFT is cut off, nor proved that such a mechanism
could be described by QFT. We suspect that the physics behind our
argument is not unconnected to the constraints on the number of
particles species that come from string theory or that appear to
be implied by entropy considerations and holography. On the other
hand, we have obtained our result without any reference to strings
or entropy, and that raises the intriguing possibility that such
implications of string theory may be more general than string
theory itself.

The QFT cutoff $\Lambda$, one expects, somehow determines the area of a
``single information bit" $A_{SIB} \sim 1/\Lambda ^2$. We may now ask
whether the size of $A_{SIB}$ given by condition (\ref{lambda}) is
compatible with the proposed statistical explanation of BH
entropy \cite{BekHaw} as given entirely by entanglement entropy
\cite{Srednicki}. Recall that the entropy of a BH is
proportional to its horizon area $A_H$ in units of Newton's
constant, $S_{BH}=A_H/4 G_N$ and does not depend on \N, while
entanglement entropy $S_{EN}={\cal N} A/A_{SIB}$ depends linearly
on \N. Considering $S_{BH}$ and $S_{EN}$ together in a way as to
make them compatible without any bound on \N, would suggest that
$A_{SIB}$ should be proportional to \N. However, condition
(\ref{lambda}) suggests that the cutoff area $1/\Lambda ^2$
scales only as ${\cal N}^{1/2}$. Thus, if the QFT cutoff
determines the true size of a single information bit, we are left
with an upper bound on \N . This upper bound,    however,  is not as
strong
as what one would obtain \cite{bgrg} by consideration of the entropy
of a BH at the ``naive" cutoff, namely the Planck scale, which
admits \N  only somewhat greater than unity.

\acknowledgements
We acknowledge helpful conversations with
R.~Bousso and J.~Friedman, and comments on the manuscript by J.
Bekenstein. R.B. and S.F. are supported by the Israel Science
foundation, S.F. is also supported  by Della Riccia Foundation
and by the Kreitman foundation. D.E. is supported by the Israel
Science Foundation, acknowledges the hospitality of the Institute
of Theoretical Physics during completion of this paper, and the
support of the National Science Foundation under Grant No.
PHY94-07194.

\end{document}